\documentclass{webofc}

\usepackage[varg]{txfonts}   
\usepackage{hyperref}
\usepackage{url}
\hypersetup{colorlinks=true,citecolor=blue,urlcolor=blue,linkcolor=blue}
\usepackage{graphbox}
\usepackage{amsmath}
\usepackage{xspace}
\usepackage{upgreek}

\usepackage{lineno}

\newcommand{\sphenix}{\mbox{sPHENIX}\xspace}
\newcommand{\vtxZ}{\mbox{$zvtx$}\xspace}
\def\mum{\ensuremath{\upmu\mathrm{m}}\xspace}
\begin{document}
\title{Intermediate Silicon Tracker in \sphenix at RHIC}
%
%

\author{\firstname{Cheng-Wei} \lastname{Shih}\inst{1,2}\fnsep\thanks{\email{cwshih0812@gmail.com}}, for the \sphenix Collaboration}

\institute{Department of Physics, and Center for High Energy and High Field Physics, National Central University, Taoyuan, Taiwan
\and 
    Nishina Center for Accelerator-Based Science, RIKEN, Wako, Japan
}

\abstract{The \sphenix collaboration has been taking data since 2023 at the Relativistic Heavy Ion Collider in BNL to study the Quark-Gluon Plasma and cold Quantum Chromodynamics (QCD). The tracking system of \sphenix consists of a time projection chamber, a MAPS-based vertex detector, and an intermediate silicon tracker (INTT). Together with the sPHENIX full barrel calorimeter system, the measurement of the heavy-flavor jets and the upsilon-state identification are enabled. INTT, surrounding the collision point azimuthally at approximately 10 cm away with two layers of silicon strip sensors, detects hit points in the intermediate area of the tracking system to enhance tracking precision. Thanks to the good timing resolution of INTT, it also provides timing information to corresponding hits of other tracking detectors. This capability eliminates pile-up events due to misidentifying bunch crossings. This proceeding discusses the achievements of INTT using Au+Au collision data taken in 2023, and the status of INTT commissioning with proton+proton collisions in 2024.
}
\maketitle
\section{Introduction}
\label{intro}
\begin{figure}[hbt!]
    \centering
    \includegraphics[width=0.45\linewidth, align = c]{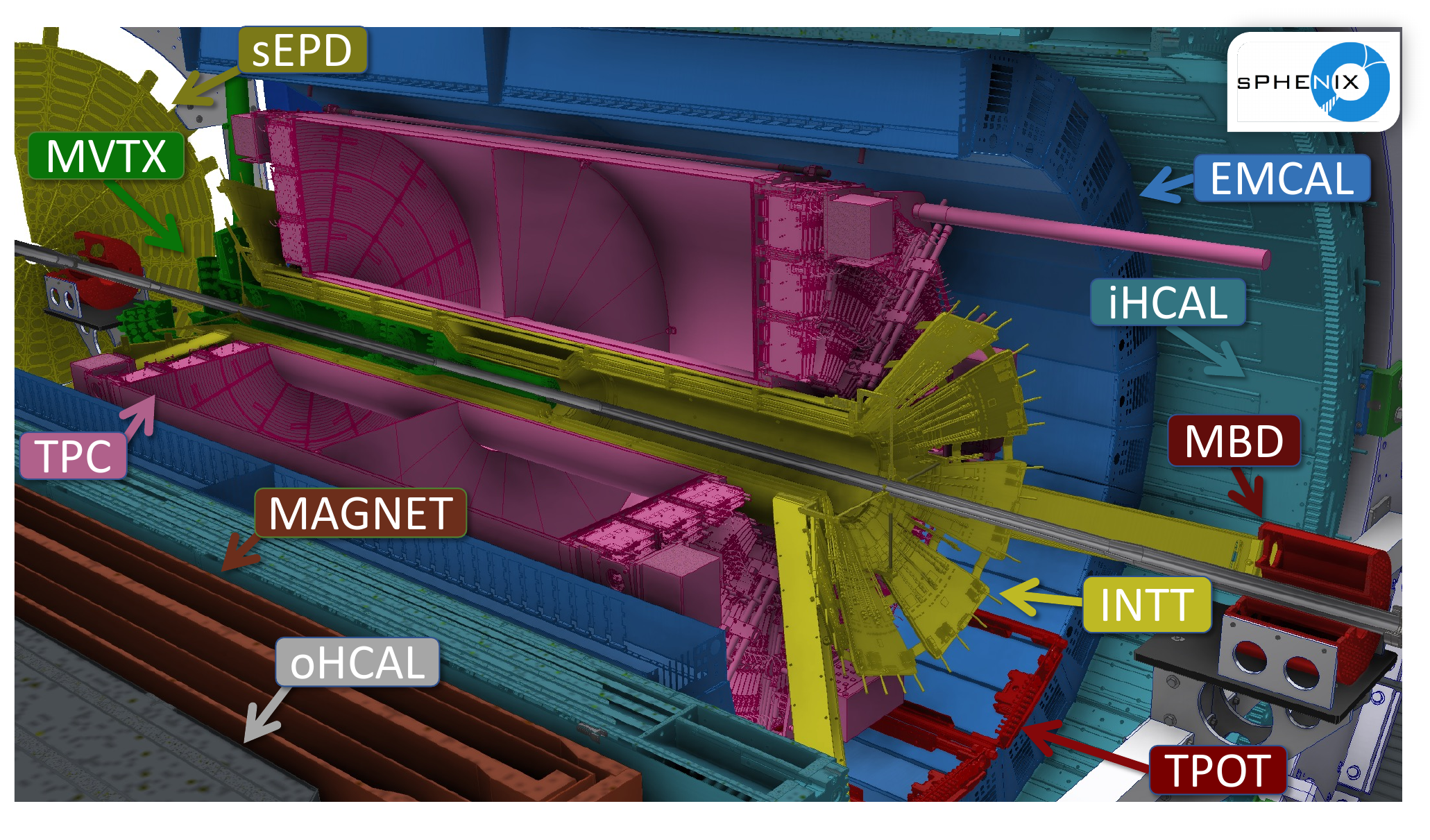}~
    \includegraphics[width=0.55\linewidth, align = c]{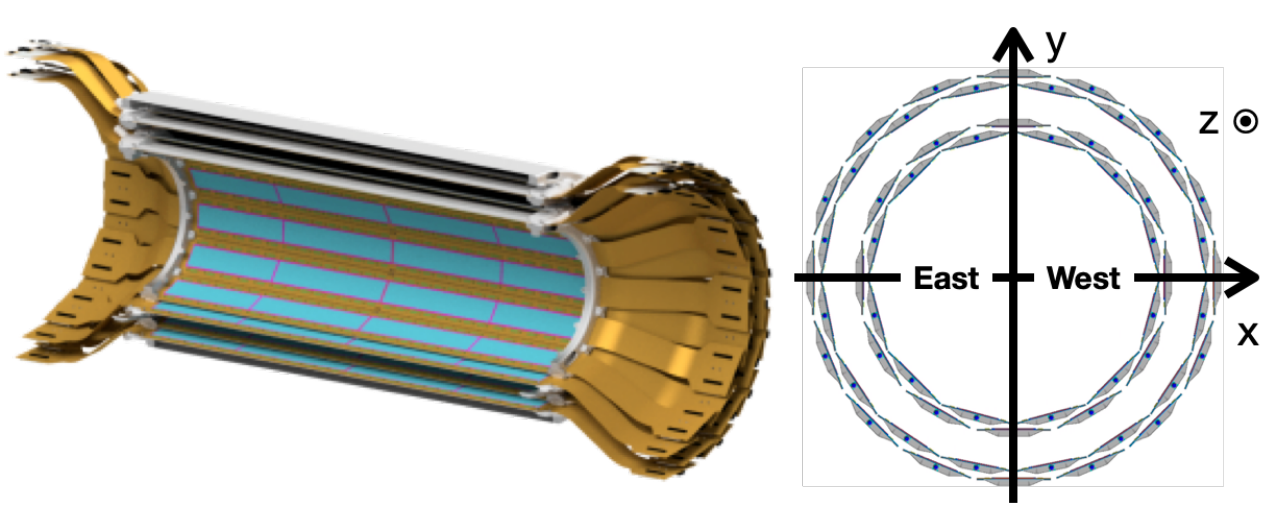}
    \caption{Left: Rendering of the sPHENIX detector. Middle and Right: Schematic and cross-section view of INTT, respectively.}
    \label{fig:detector_plot}
\end{figure}

\sphenix, a new collider experiment built after 20-year operation of the Relativistic Heavy Ion Collider (RHIC) at Brookhaven National Laboratory, is a state-of-the-art jet detector aiming to study the strongly interacting Quark-Gluon Plasma and cold Quantum Chromodynamics by measuring jets, jet correlations, and leptons with high precision~\cite{Ref_sPH_TDR}. Figure~\ref{fig:detector_plot} shows the cutaway rendering of the \sphenix detector. Driven by these goals, the \sphenix detector features large acceptance, full barrel calorimeters, a 1.4 Tesla solenoid magnet, and an advanced tracking system. The full barrel calorimeter system, first time available at RHIC, stands for the electromagnetic calorimeter (EMCal), and inner and outer hadronic calorimeters (iHCal and oHCal) covering full azimuthal angle and pseudorapidity $|\eta| < 1.1$. The tracking system of \sphenix is composed of a monolithic active pixel sensor vertex detector (MVTX) for the vertex measurement, an intermediate silicon tracker (INTT) for the track timing information, a time projection chamber (TPC) for the track momentum measurement, and a TPC outer tracker (TPOT) for the TPC-distortion calibration. 

INTT is a two-layer barrel strip tracker located in the intermediate region of the \sphenix tracking system. INTT offers two spatial points to bridge tracks of the MVTX and TPC, and enhance the pattern recognition~\cite{CW_INTT}. In addition, the timing resolution of INTT, which is finer than the bunch-crossing frequency of RHIC and is the best in the tracking system, enables INTT to associate reconstructed tracks with individual bunch crossings, eliminating pile-up events. The mechanical drawing of INTT is shown in Figure~\ref{fig:detector_plot}. The INTT strips are 78 \mum in pitch width, 20 or 16 mm in length,
which give INTT a good spatial resolution in polar angle, while the resolution in determining the hit position in the beam direction (the z-axis) is relatively coarse.

\section{INTT performance in Au+Au collisions in year 2023}
\label{Run_2023}
\begin{figure}[hbt!]
    \centering
    \includegraphics[width=0.5\linewidth, align = c]{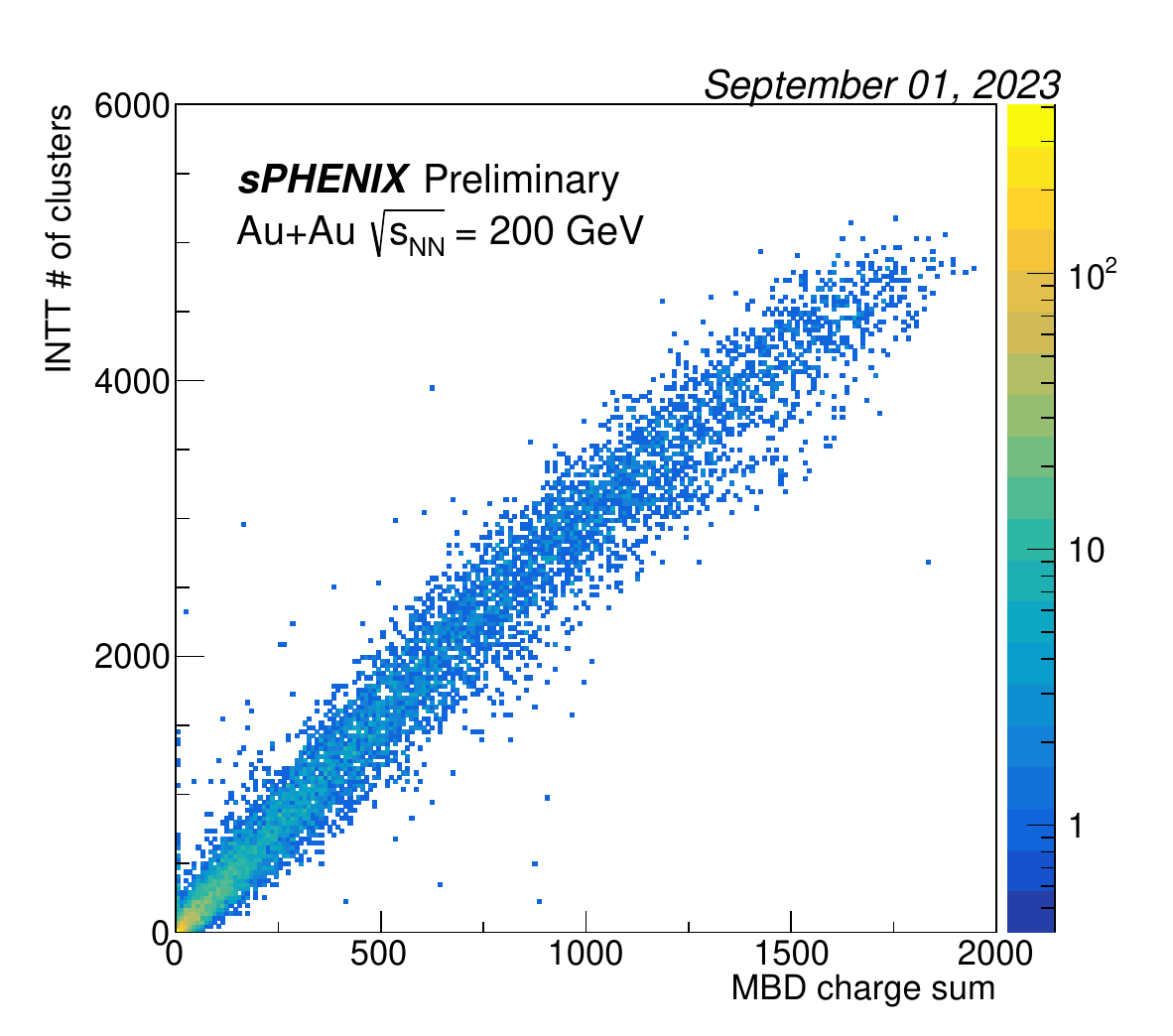}~
    \includegraphics[width=0.5\linewidth, align = c]{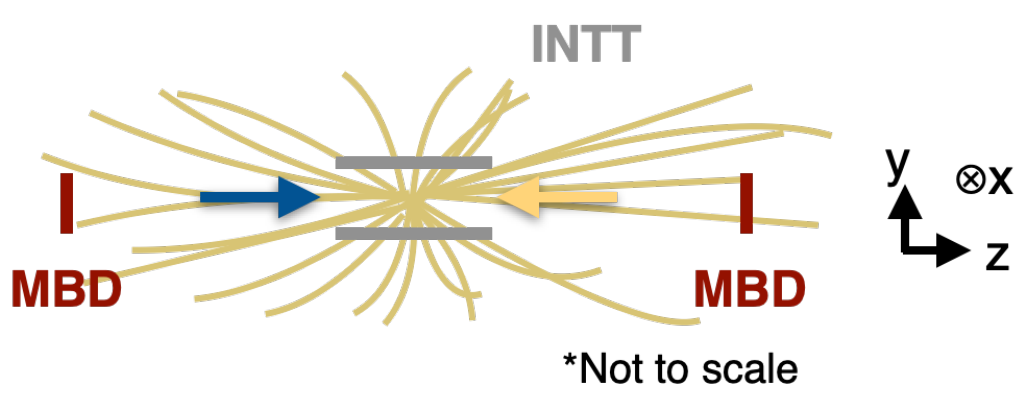}
    \caption{Left: Correlation between the MBD total charge and the INTT number of clusters. Right: Illustrating the relative locations of INTT and MBD.}
    \label{fig:INTT_multi_corr_trigger}
\end{figure}

After more than 10 years of preparation, \sphenix started to take data with Au+Au collisions in May 2023~\cite{Brien_sPH}. INTT was commissioned in the triggered-readout mode shortly after. Figure~\ref{fig:INTT_multi_corr_trigger} shows the correlation between the measured multiplicity of the Minimum Bias Detector (MBD) of sPHENIX and that of INTT. The positive correlation is identified, indicating that the INTT and MBD are timed in, and the recorded data in both subsystems are reliable. 

\begin{figure}[hbt!]
    \centering
    \includegraphics[width=1.0\linewidth]{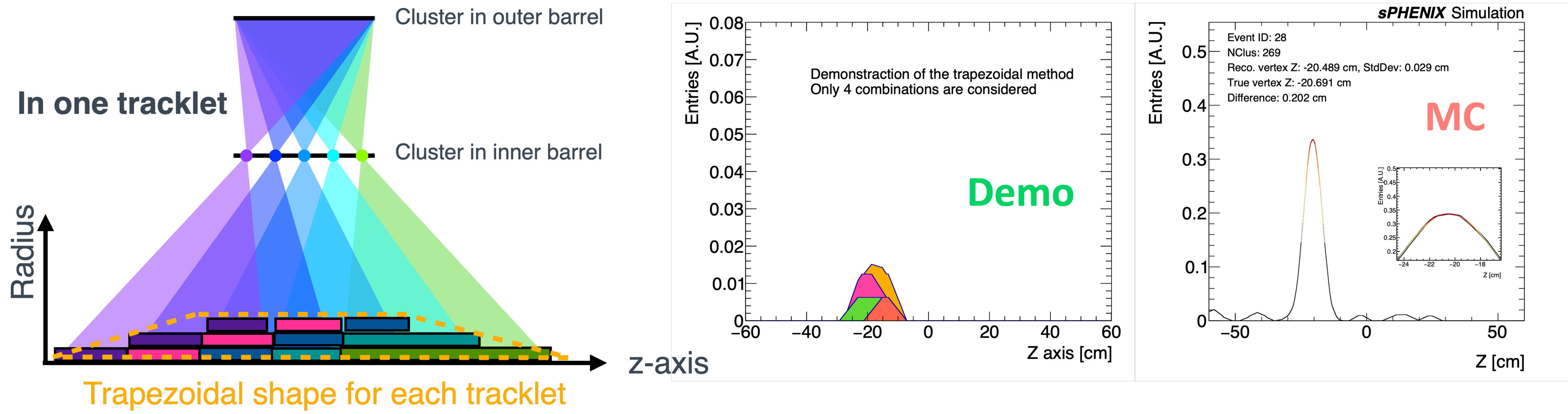}
    \caption{Left: Demonstration of the probability distribution of collision location in the z-axis inferred from one tracklet. Middle and Right: The Gaussian-like distribution formed by stacking up the trapezoidal shapes of the tracklets in demonstration and simulation, respectively.}
    \label{fig:trapezoidal}
\end{figure}

\begin{figure}[hbt!]
    \centering
    \includegraphics[width=0.5\linewidth]{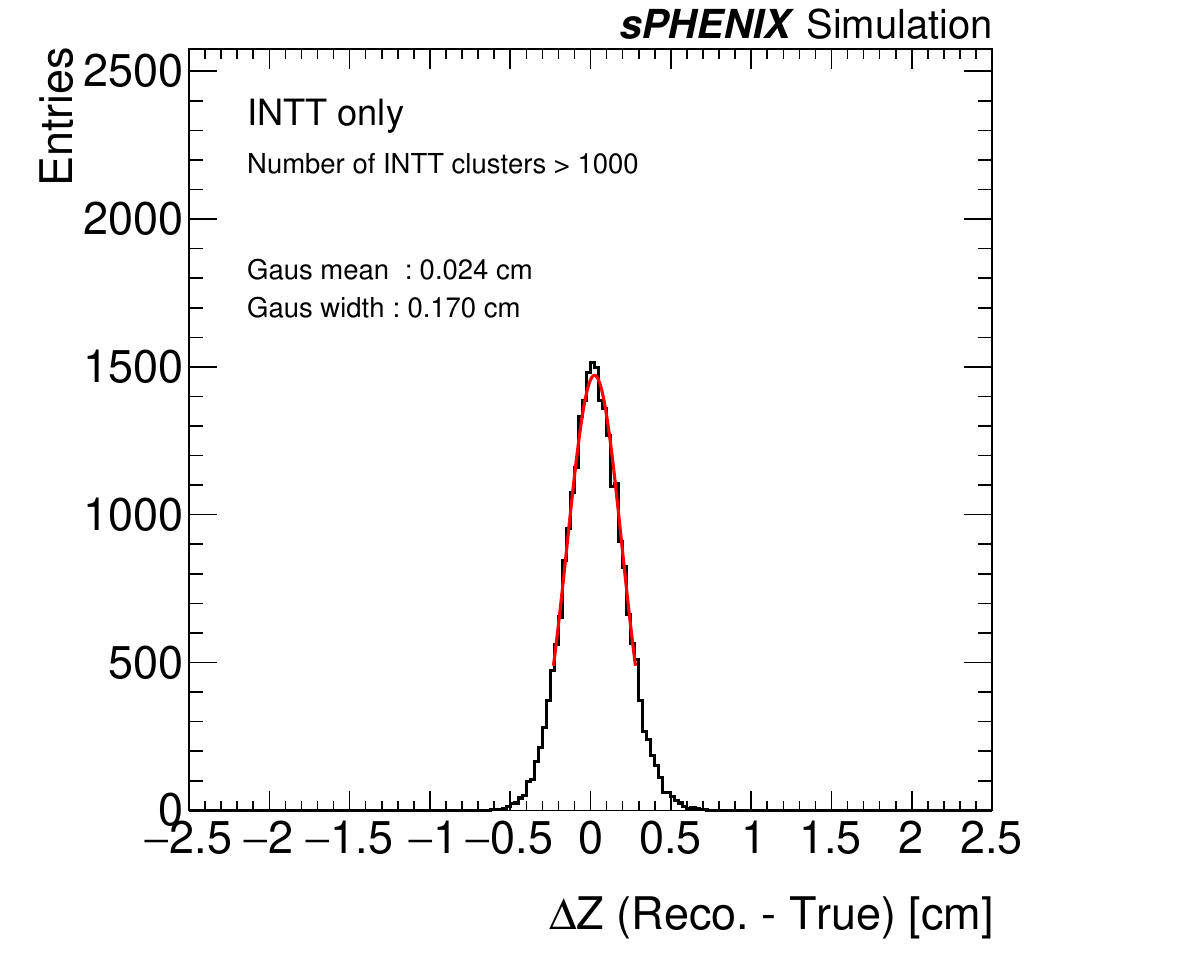}~
    \includegraphics[width=0.5\linewidth]{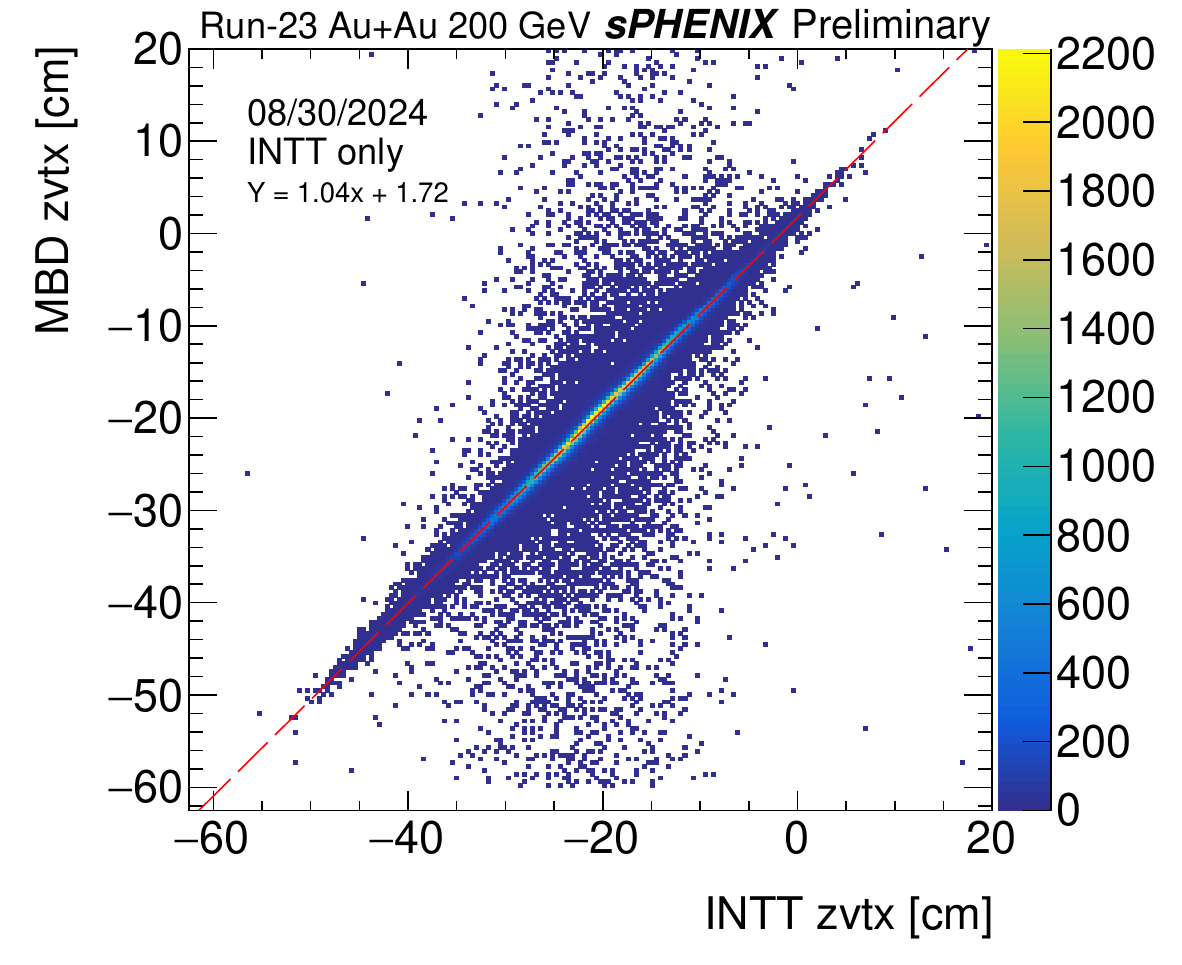}
    \caption{Left: Distribution of difference between the reconstructed INTT \vtxZ and truth \vtxZ in simulation. Right: Correlation between the z-vertices reconstructed by INTT and MBD in data.}
    \label{fig:trapezoidal_performance}
\end{figure}

Additionally, comprehensive analyses were conducted to enhance the understanding of INTT, such as the z-vertex (\vtxZ) reconstruction. A strip tracker is not conventionally designed for reconstructing the collision vertex, as its strip lengths, on the order of centimeters, may be coarse for this purpose. The potential of which by INTT was investigated since it could potentially provide a valuable reference for calibrating the z-vertex measured by MBD. The procedures of the reconstruction algorithm developed for INTT are briefly described as follows. Inner-barrel clusters are paired with outer-barrel clusters. Pairs with small differences in azimuthal angle are retained as tracklets.
Since a particle can interact with an INTT strip at any position along its length in the beam direction (the z-axis), the probability distribution of collision location inferred from a tracklet is in the shape of a trapezoid, as demonstrated in Figure~\ref{fig:trapezoidal}. By stacking the trapezoidal shapes formed by all the tracklets in a single event, a Gaussian-like distribution is formed, as the example shown in Figure~\ref{fig:trapezoidal}. The \vtxZ of an event is determined by the average fit Gaussian means. The performance of this approach with INTT was evaluated in simulation, as shown in Figure~\ref{fig:trapezoidal_performance}. The resolution can reach up to 1.7 mm in events with the number of clusters exceeding one thousand, which is 10 times smaller than the intrinsic strip length of INTT. The algorithm was also applied to data, and the \vtxZ measured by MBD and INTT were compared, as shown in Figure~\ref{fig:trapezoidal_performance}. The positive correlation with the slope close to unity validates the algorithm and provides additional confirmation of the integrity of the INTT recorded data.

\section{INTT performance in proton+proton collisions in year 2024}
\label{Run_2024}
\begin{figure}[hbt!]
    \centering
    \includegraphics[width=0.47\linewidth]{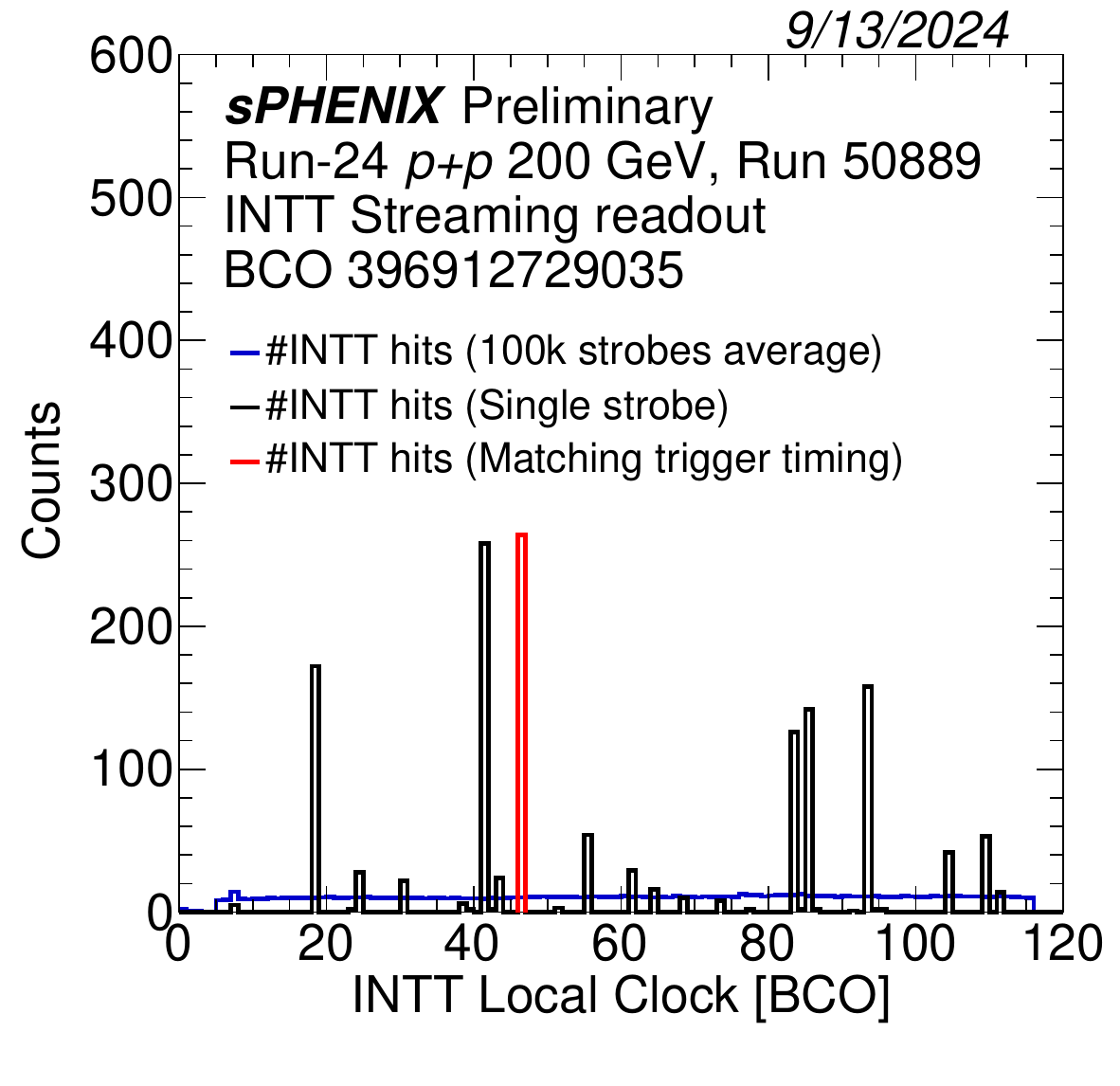}~
    \includegraphics[width=0.50\linewidth]{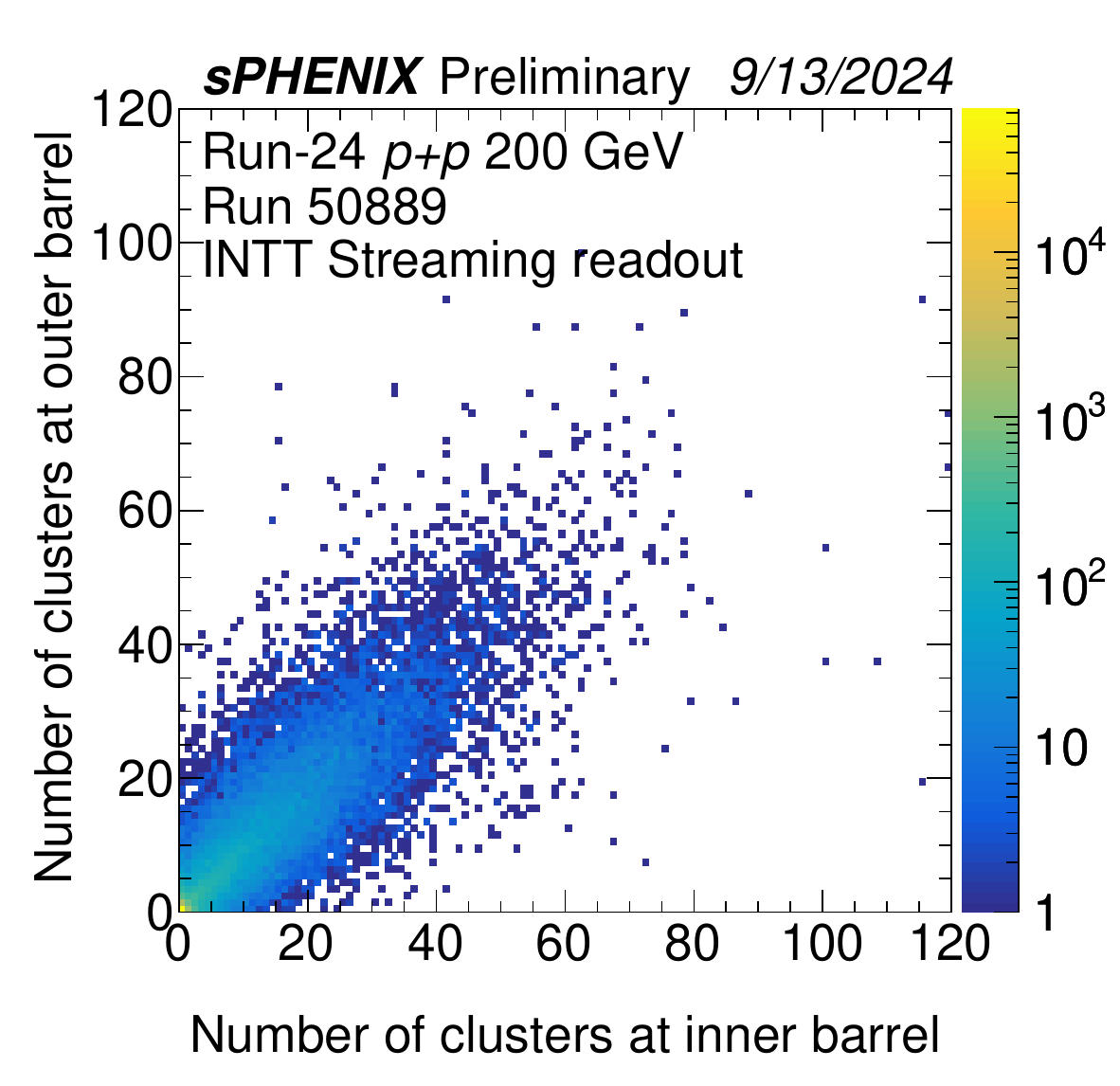}
    \caption{Left: INTT hit counts as a function of bunch crossing index in a single time frame. Right: Correlation between the INTT numbers of inner-barrel clusters and outer-barrel clusters.}
    \label{fig:stream_corelation}
\end{figure}

In 2024, \sphenix started to collect the proton+proton collision data in April. During the data taking, the INTT data acquisition mode was transitioned into a streaming readout mode. Unlike the triggered-readout mode, which keeps events only if triggers are fired, such as the EMCal photon trigger, the streaming readout mode enables recording events as long as the particle hits are detected by INTT. The streaming readout is crucial for the sPHENIX heavy-flavor physics program as the objects of interest are not energetic enough to fire the triggers~\cite{Cameron_HF}. Figure~\ref{fig:stream_corelation} shows the number of INTT hit counts as a function of bunch crossing (BCO) index. The bin in red indicates the collected hits associated with a trigger, while the spikes in black represent the recorded hits from the rest of the collision events without a trigger fired. The capability of INTT operating in streaming readout mode is demonstrated. Besides, the cluster multiplicities of two INTT barrels were correlated as a sanity check, as shown in Figure~\ref{fig:stream_corelation}. The observed positive correlation suggests that the INTT operation in streaming readout mode is functioning as expected.

\begin{figure}[hbt!]
    \centering
    \includegraphics[width=1.0\linewidth]{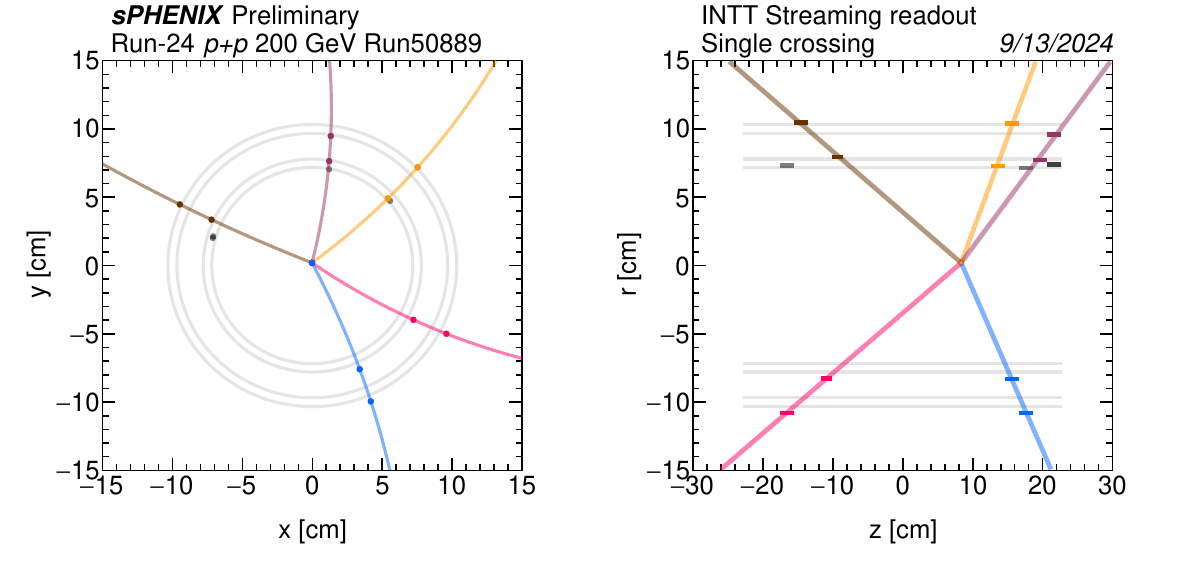}
    \caption{Event display of INTT performing track reconstruction with INTT clusters in the x-y plane (Left) and the z-radius plane (Right).}
    \label{fig:stream_tracking}
\end{figure}

\begin{figure}[hbt!]
    \centering
    \includegraphics[width=0.4\linewidth, align = c]{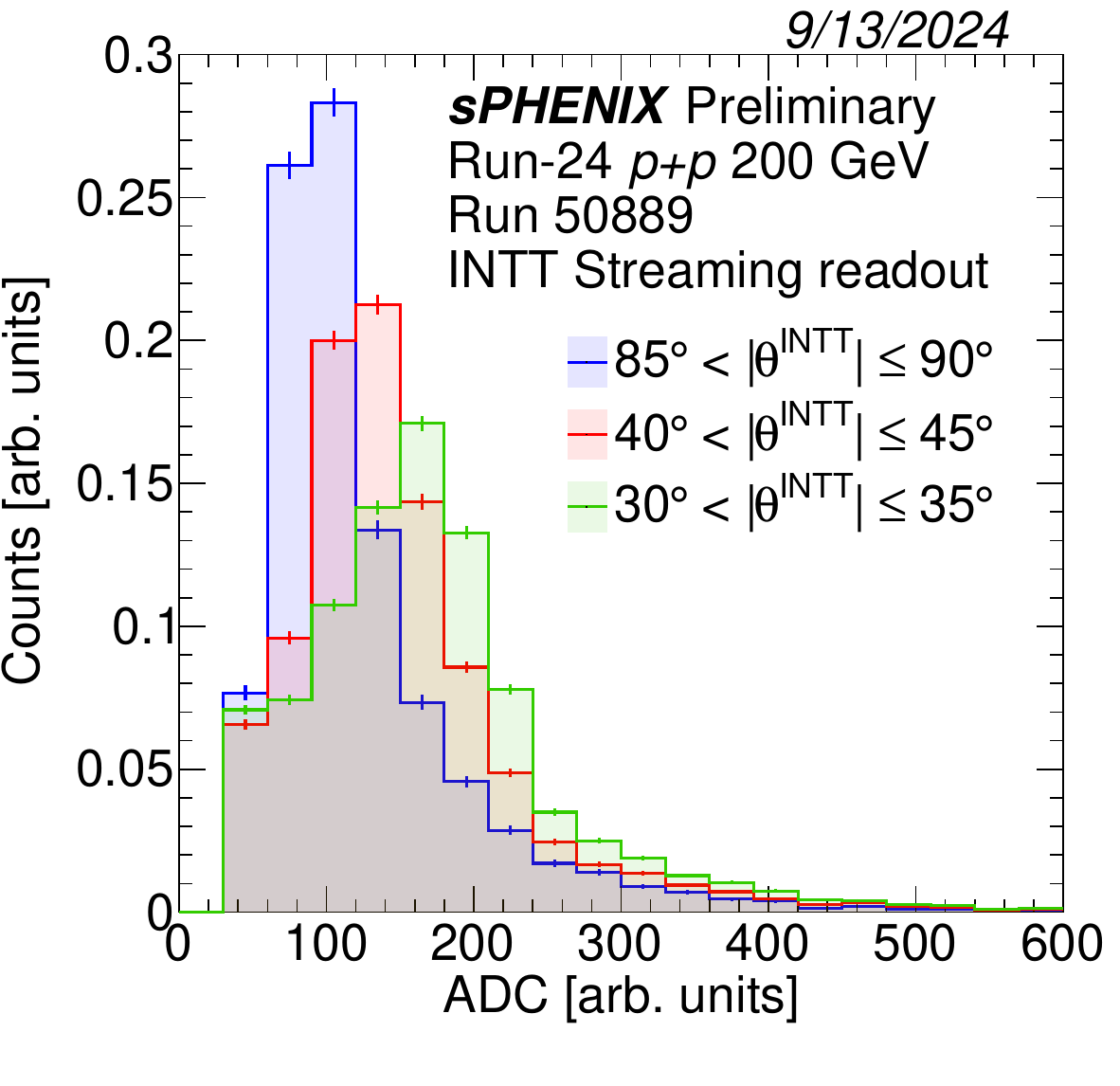}~
    \includegraphics[width=0.5\linewidth, align = c]{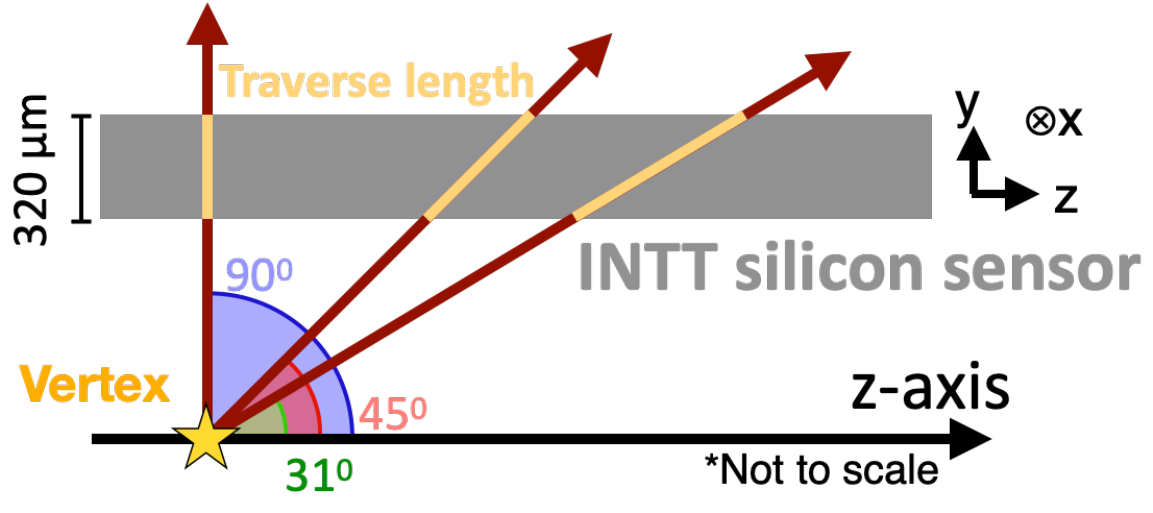}\\
    \caption{Left: Particle energy deposit distributions with different particle incident-angle intervals. Right: Illustrating the relation between particle incident angle and traverse length in the INTT sensor.}
    \label{fig:stream_edep}
\end{figure}

The detailed analyses were further performed with the aim of gaining more insights into the recorded streaming data, such as the INTT standalone track reconstruction. An event display is shown in Figure~\ref{fig:stream_tracking}. Particle tracks are reconstructed by utilizing the INTT clusters on the top of the vertex reconstruction. The bending curvatures due to the magnetic field are clearly identified. The mission as a tracker is proven to be achievable. Furthermore, an evident incident-angle dependence was observed in the study of particle energy deposition in the INTT sensors, as shown in Figure~\ref{fig:stream_edep}. A smaller particle incident angle leads to a longer traverse length, resulting in a larger amount of energy deposited in the INTT sensor. The measured result, exhibiting a consistent trend, confirms that the recorded data predominantly consists of signal hits.

\section{Conclusion}
\label{conclusion}
The timing resolution of INTT, which is finer than the bunch-crossing frequency of RHIC, makes it the only tracking detector in sPHENIX capable of associating individual tracks with events. In Run 2023, sPHENIX started commissioning with Au+Au collisions, during which a correlation in multiplicity between INTT and MBD was identified. Besides, the z-vertices reconstructed by INTT and MBD showed a positive correlation with a slope close to unity. During Run 2024, when sPHENIX collected proton+proton collision data, INTT transitioned to the streaming readout mode. This is crucial for heavy-flavor physics as all collision events can be recorded. In streaming readout mode, a clear multiplicity correlation was observed between the INTT inner and outer barrels. In addition, the developed INTT tracklet analysis was able to reconstruct particle tracks. Moreover, the distinct particle incident-angle dependence of energy deposition in the INTT sensor was concluded. INTT has been confirmed to be in good condition and reliable in both Run 2023 and Run 2024! With the substantial statistics taken, sPHENIX is going to deliver exciting physics results!


%
%
%

\end{document}